\documentstyle[preprint,aps,epsfig,tighten]{revtex}	
\newcommand{\ppbar}{\mbox{$p\overline{p}$}}
\newcommand{\bbbar}{\mbox{$b\overline{b}$}}
\newcommand{\ccbar}{\mbox{$c\overline{c}$}}
\newcommand{\ttbar}{\mbox{$t\overline{t}$}}
\newcommand{\ee}{\mbox{$ee$}}
\newcommand{\etal}{{\it et al.}}
\newcommand{\met}{\mbox{${E\!\!\!\!/_T}$}}
\newcommand{\rpv}{\mbox{${R\!\!\!\!\!\:/_p}$}}
\newcommand{\rp}{\mbox{${R_p}$}}
\newcommand{\lamp}{\mbox{$\lambda_{121}'$}}
\newcommand{\intlum}{\mbox{${ \int {\cal L} \; dt}$}}
\newcommand{\pt}{\mbox{$p_T$}}
\newcommand{\et}{\mbox{$E_T$}}
\newcommand{ \qbar}     {\mbox{$\overline{q}$}}
\newcommand{ \gluino}   {\mbox{$\tilde{g}$}}
\newcommand{ \squark}   {\mbox{$\tilde{q}$}}

\newcommand{ \mgluino}  {\mbox{$M(\gluino)$}}
\newcommand{ \msquark}  {\mbox{$M(\squark)$}}
\newcommand{ \csquarkl} {\mbox{$\tilde{c}_L$}}
\newcommand{ \mcsl}     {\mbox{$M(\csquarkl)$}}
\newcommand{ \ssb}     {\mbox{$\squark\overline{\squark}$}}

\newcommand{ \tsquark} {\mbox{$\tilde{t}$}}
\newcommand{ \stopo}   {\mbox{$\tilde{t}_1$}}

\newcommand{ \ttbone}  {\mbox{$\tsquark_1\overline{\tsquark}_1$}}
\newcommand{ \chizero}{\mbox{$\tilde{\chi}_{1}^0$}}
\newcommand{ \mchio}{\mbox{$M(\chizero)$}}
\newcommand{ \chione }{\mbox{$\tilde{\chi}_{1}^{\pm}$}}
\newcommand{ \mchione }{\mbox{$M(\tilde{\chi}_{1}^{\pm})$}}
\newcommand{ \mntwo  }{\mbox{$M(\tilde{\chi}_{2}^{0})$}}
\newcommand{ \mstopo}   {\mbox{$M(\tilde{t}_1)$}}
\newcommand{\gev}  {\mbox{$\;{\rm GeV}$}}

\newcommand{\gevc} {\mbox{$\;{\rm GeV}/c$}}
\newcommand{\gevcc}{\mbox{$\;{\rm GeV}/c^2$}}
\newcommand{\tevcc}{\mbox{$\;{\rm TeV}/c^2$}}
\newcommand{\ipb}{\mbox{${\rm pb}^{-1}$}}

\newcommand{\pelp}{\mbox{$e^+$}}

\newcommand{\pelpm}{\mbox{$e^{\pm}$}}
\newcommand{\epem}{\mbox{$e^+e^-$}}
\newcommand{\ifb}{\mbox{${\rm fb}^{-1}$}}

\begin{document}
\draft
\title{
\boldmath
Search for $R$-parity Violating Supersymmetry using Like-Sign Dielectrons 
in \ppbar\ Collisions at $\sqrt{s} = 1.8$ TeV}
\maketitle
\font\eightit=cmti8
\def\r#1{\ignorespaces $^{#1}$}
\hfilneg
\begin{sloppypar}
\noindent
F.~Abe,\r {17} H.~Akimoto,\r {39}
A.~Akopian,\r {31} M.~G.~Albrow,\r 7 A.~Amadon,\r 5 S.~R.~Amendolia,\r {27} 
D.~Amidei,\r {20} J.~Antos,\r {33} S.~Aota,\r {37}
G.~Apollinari,\r {31}
% ...
\end{sloppypar}
\vskip .026in
\begin{center}
(CDF Collaboration)
\end{center}

\vskip .026in
\begin{center}
\r 1  {\eightit Argonne National Laboratory, Argonne, Illinois 60439} \\
\r 2  {\eightit Istituto Nazionale di Fisica Nucleare, University of Bologna,
I-40127 Bologna, Italy} \\
\r 3  {\eightit Brandeis University, Waltham, Massachusetts 02254} \\
\r 4  {\eightit University of California at Los Angeles, Los 
Angeles, California  90024} \\  
\r 5  {\eightit University of Chicago, Chicago, Illinois 60637} \\
\r 6  {\eightit Duke University, Durham, North Carolina  27708} \\
\r 7  {\eightit Fermi National Accelerator Laboratory, Batavia, Illinois 
60510} \\
\r 8  {\eightit University of Florida, Gainesville, Florida  32611} \\
\r 9  {\eightit Laboratori Nazionali di Frascati, Istituto Nazionale di Fisica
               Nucleare, I-00044 Frascati, Italy} \\
\r {10} {\eightit University of Geneva, CH-1211 Geneva 4, Switzerland} \\
\r {11} {\eightit Harvard University, Cambridge, Massachusetts 02138} \\
\r {12} {\eightit Hiroshima University, Higashi-Hiroshima 724, Japan} \\
\r {13} {\eightit University of Illinois, Urbana, Illinois 61801} \\
\r {14} {\eightit Institute of Particle Physics, McGill University, Montreal 
H3A 2T8, and University of Toronto,\\ Toronto M5S 1A7, Canada} \\
\r {15} {\eightit The Johns Hopkins University, Baltimore, Maryland 21218} \\
\r {16} {\eightit Institut f\"{u}r Experimentelle Kernphysik, 
Universit\"{a}t Karlsruhe, 76128 Karlsruhe, Germany} \\
\r {17} {\eightit National Laboratory for High Energy Physics (KEK), Tsukuba, 
Ibaraki 305, Japan} \\
\r {18} {\eightit Ernest Orlando Lawrence Berkeley National Laboratory, 
Berkeley, California 94720} \\
\r {19} {\eightit Massachusetts Institute of Technology, Cambridge,
Massachusetts  02139} \\   
\r {20} {\eightit University of Michigan, Ann Arbor, Michigan 48109} \\
\r {21} {\eightit Michigan State University, East Lansing, Michigan  48824} \\
\r {22} {\eightit University of New Mexico, Albuquerque, New Mexico 87131} \\
\r {23} {\eightit The Ohio State University, Columbus, Ohio  43210} \\
\r {24} {\eightit Osaka City University, Osaka 588, Japan} \\
\r {25} {\eightit Universita di Padova, Istituto Nazionale di Fisica 
          Nucleare, Sezione di Padova, I-35131 Padova, Italy} \\
\r {26} {\eightit University of Pennsylvania, Philadelphia, 
        Pennsylvania 19104} \\   
\r {27} {\eightit Istituto Nazionale di Fisica Nucleare, University and Scuola
               Normale Superiore of Pisa, I-56100 Pisa, Italy} \\
\r {28} {\eightit University of Pittsburgh, Pittsburgh, Pennsylvania 15260} \\
\r {29} {\eightit Purdue University, West Lafayette, Indiana 47907} \\
\r {30} {\eightit University of Rochester, Rochester, New York 14627} \\
\r {31} {\eightit Rockefeller University, New York, New York 10021} \\
\r {32} {\eightit Rutgers University, Piscataway, New Jersey 08855} \\
\r {33} {\eightit Academia Sinica, Taipei, Taiwan 11530, Republic of China} \\
\r {34} {\eightit Texas A\&M University, College Station, Texas 77843} \\
\r {35} {\eightit Texas Tech University, Lubbock, Texas 79409} \\
\r {36} {\eightit Istituto Nazionale di Fisica Nucleare, University of Trieste/
Udine, Italy} \\
\r {37} {\eightit University of Tsukuba, Tsukuba, Ibaraki 315, Japan} \\
\r {38} {\eightit Tufts University, Medford, Massachusetts 02155} \\
\r {39} {\eightit Waseda University, Tokyo 169, Japan} \\
\r {40} {\eightit University of Wisconsin, Madison, Wisconsin 53706} \\
\r {41} {\eightit Yale University, New Haven, Connecticut 06520} \\
\end{center}
\vspace{-.3in}
\begin{abstract}
We present a search for like-sign dielectron plus multijet events 
using 107 \ipb\ of data in \ppbar\ collisions at $\sqrt{s} = 1.8$ TeV 
collected in 1992-95 by the CDF experiment.  
Finding no events that pass our selection, we set $\sigma\cdot Br$ limits on
two SUSY processes that can produce this experimental signature:
gluino-gluino or squark-antisquark production
with $R$-parity violating decays of the charm squark 
or lightest neutralino via a non-zero \lamp\ coupling.  
We compare our results to NLO
calculations for gluino and squark production cross sections
and set lower limits on \mgluino, \mstopo, and \msquark.    
\end{abstract}
\vspace{.2in}
\vspace{.2in}
\narrowtext
The minimal supersymmetric standard model (MSSM) \cite{mssm} is an
extension of the standard model (SM) that adds a supersymmetric (SUSY)
partner for each SM particle and is constructed
to conserve baryon number ($B$) and lepton number ($L$).  The
requirement of $R$-parity (\rp) \cite{rparity}
conservation is imposed on the couplings: for a particle of spin $S$, the
multiplicative quantum number $\rp \equiv (-1)^{3B+L+2S}$ 
distinguishes SM particles ($\rp = +1$) from SUSY particles ($\rp = -1$).  If
\rp\ is conserved, SUSY particles can only be produced in pairs and the
lightest supersymmetric particle (LSP) is stable.  The assumption 
of \rp\ conservation thus leads to experimental signatures with appreciable 
missing transverse energy (\met), provided that the LSP is electrically  
neutral and colorless \cite{ellis}.
\rp\ conservation, however, is not required by SUSY theories in general and
viable $R_p$ violating (\rpv) models can be built by adding explicitly 
$B$ {\em or} $L$ violating couplings to the SUSY Lagrangian \cite{rpvtheory}.
Since the LSP can be unstable in this case, the 
standard \met\ signature is diluted.

The results at high $Q^2$ from the HERA experiments \cite{hera} have sparked
interest in \rpv\ SUSY, since
the excess of events observed at high $Q^2$ could be explained by
the production and decay of a single squark: $\pelp 
+ d \to \squark \to \pelp+ d$, where \rp\ is violated at
both vertices \cite{choud,choud_prd,drei_mora,altarelli}.  
In this scenario, $\csquarkl$ (the SUSY partner of the left-handed
charm quark) with mass
$\mcsl\simeq200\gevcc$ is the preferred squark 
flavor because its associated \rpv\ Yukawa coupling \lamp\ 
is less constrained by experiment than the other couplings \cite{bounds}.
Another possibility to explain the excess is the production and decay
of a first-generation leptoquark; D\O\ and CDF have
ruled out this explanation \cite{dzerocdflq}.

In this Letter, we examine two \rpv\ processes in an MSSM framework
that involve the same \lamp\ coupling: (1) $\ppbar\to\gluino\gluino \to
(c\,\csquarkl)\,(c\,\csquarkl) \to
c\,(\pelpm d)\,c\,(\pelpm d)$ ``charm squark analysis''; and (2)
$\ppbar\to\ssb \to (q\chizero)\,(\bar{q}\chizero) 
\to q\,(dc\pelpm)\, \bar{q}\,(dc\pelpm)$ ``neutralino analysis''.
For process (1) we assume $\msquark > \mgluino > \mcsl=200 \gevcc$,
where $\msquark$ denotes the degenerate mass
for all up-type (except for \csquarkl) 
and all right-handed down-type squarks.  The masses of the 
left-handed down-type squarks are calculated using the relations given in 
Reference~\cite{choud}.  These assumptions are motivated by the HERA results.  
Process (2) is a complementary search also based on $\lamp\neq0$.
It is favored if the size of the
\rpv\ coupling is small compared to the SM gauge couplings.
We separately consider \ssb\ production (5 degenerate squark flavors) 
and \ttbone\ production, and make the mass assumptions:
$M(\chione),\mntwo\,>\, \msquark\,>\,\mchio$, where \squark\ refers here to 
either the degenerate squark or \stopo, and $M(\chione) \approx 2\,\mchio$.
The first relation suppresses $\squark \to \chione\,+\,X$ and the
second approximation is generally true for most
combinations of SUSY parameters, particularly when assumptions leading
to gaugino mass unification are made.
For the case of \ttbone\ production, we further assume 
$\mchione>\mstopo-M(b)$
to ensure that $Br(\stopo \to c\chizero)=100\%$ for the relevant case:
$\mstopo<M(t)$.
For these two searches, we make the conservative and simplifying 
assumption that there is only one non-zero \rpv\ coupling.
Given the Majorana nature of the gluino and neutralino, reactions (1)
and (2) each yield
like-sign (LS) and opposite-sign (OS) dielectrons with equal
probability.  Since
LS dilepton events have the benefit of small SM backgrounds,
we search for events with LS dielectrons and two or more jets.

We present results of a search for
$\ppbar \to e^{\pm}e^{\pm} + \geq 2$ jet events using 107 \ipb\ of data
from \ppbar\ collisions at a center of mass energy of $\sqrt{s}$ = 1.8 TeV.
The data were collected by the Collider Detector at 
Fermilab (CDF) \cite{det} during the 1992-93 and 1994-95 runs 
of the Fermilab Tevatron.
At CDF the location of the \ppbar\ collision event vertex ($z_{vertex}$)
is measured along the beam direction with a time projection chamber. 
The transverse momenta (\pt) of charged particles are measured 
in the pseudorapidity region $|\eta|<$ 1.1 with a
drift chamber, which is located in a 1.4~T solenoidal magnetic field.
Here $\pt = p \sin \theta$ and $\eta$ = $\rm - \ln \tan(\theta/2)$, where
$\theta$ is the polar angle with respect to the proton beam direction.
The electromagnetic (EM) and hadronic calorimeters are segmented in a 
projective tower geometry surrounding the solenoid
and cover the central ($|\eta|<1.1$) and plug ($1.1<|\eta|<2.4$) regions.  
A gas proportional chamber located at shower maximum in the central EM
calorimeter provides shower position and profile
measurements in both the $z$ and $r-\phi$ directions.

Dielectron plus multijet candidates are selected from events
that pass the central electron triggers with 
$\et(e) > 9.2\gev$ in the 1992-93 run, while for the 1994-95 run
there are two such triggers, with thresholds of 8 and 16 \gev.  The 8 
\gev\ trigger imposes additional requirements on the development of the 
EM shower.
In our analysis, we require two electrons with $\et>15\gev$.  
Each electron candidate must exhibit a lateral shower 
profile consistent with that which is expected for electrons,
be well matched to a track \cite{wmass1a} with 
$\pt \geq \et/2$, and pass a sliding cut on 
the ratio of energy in the hadron calorimeter to the energy in the
EM calorimeter (hadronic energy fraction) \cite{top}.
At least one electron candidate must also pass more stringent
identification requirements on its shower profile and hadronic energy
fraction \cite{cdflll_1a}. 
Each electron must pass an isolation cut in which
the total calorimeter \et\ in an $\eta-\phi$ cone of radius
$R \equiv \sqrt{(\Delta\phi)^2+(\Delta\eta)^2}= 0.4$ 
around the electron, excluding the electron \et, is less than 4 GeV.
This helps to remove the background from \bbbar\ and \ccbar\ production 
(\bbbar/\ccbar) while retaining much of
the sensitivity to the SUSY signal.  The $\eta-\phi$ distance 
$\Delta R_{ee}\equiv\sqrt{(\Delta\phi_{ee})^2+(\Delta\eta_{ee})^2}$
between the two electrons must be greater than 0.4 to avoid
shower overlap in the calorimeter.
The event $|z_{vertex}|$ must be less than 60~cm to restrict the
analysis to the region of the detector that retains the projective
nature of the calorimeter towers, and
both electrons must be consistent with originating from the same vertex.
Jets are identified in the calorimeter using a fixed cone clustering
algorithm \cite{cdfjetalg} with cone size $R =0.7$.
We require at least two jets with $\et >$ 15 GeV and $|\eta_j| <
2.4$, separated by $\Delta R_{jj}>0.7$, and $\Delta R_{ej}>0.7$.
Finally, there must be no significant \met\ in the event: 
$\met/\sqrt{\sum\et} < 5\gev^{1/2}$,   
where $\sum \et$  is the scalar sum of transverse energy in the
calorimeter for the two electrons and two leading jets.
These selection requirements are effective in removing
the $\bbbar/\ccbar$ and \ttbar\ backgrounds while retaining the signal.
No LS candidate events survive this selection, while 165 OS events are 
retained.

We calculate the event acceptance using Monte Carlo samples generated
with ISAJET 7.20 \cite{isajet}, CTEQ3L parton
distribution functions \cite{cteq3l}, and passed through the CDF
detector simulation program.  
For the charm squark analysis, we examine four values of the gluino mass:
210, 250, 300, and 400 \gevcc\ while 
the charm squark mass, \mcsl, is fixed at 200\gevcc.  For the neutralino
analysis, we create Monte Carlo samples with \msquark\
in the range $100-350\gevcc$.  For each \msquark, we
generate samples for two extremes of the neutralino mass:
$\mchio=\msquark/2$, which corresponds to $\mchione\simeq\msquark$, and 
$\mchio=\msquark-M(q)$, the kinematic limit for the decay.

The dominant SM backgrounds for this search are
\ttbar\ and $\bbbar/\ccbar$ production, where both can give rise to LS
$ee$ events.
We use ISAJET 7.20 \cite{isajet} Monte Carlo
samples to estimate the sizes of these backgrounds.  
For \ttbar\ production and decay, 
we analyze 25K events (corresponding to $\intlum=3.3\,\ifb$)
with $M(t)$ = 175 \gevcc\ and
$\sigma_{\ttbar}$ = 7.6 pb \cite{cdf_tt_xs}
and find zero accepted LS $ee$ events.
Top dilepton events typically have appreciable \met\ and are rejected
by the \met\ significance cut.
We study Monte Carlo samples of \bbbar/\ccbar\ events for 
two different processes: direct production
and final state gluon splitting, and expect a contribution of
$0.3\pm0.3$ LS events from this source in 107 \ipb.  The isolation
cut on the electrons is efficient in removing this background
as semileptonic $b$ quark decays yield poorly isolated leptons.  
The total expected background is therefore consistent with zero events, so
we forego background subtraction in setting limits.  
The remaining 165 OS events are consistent with 
the expected contribution of $153.0\pm14.5$ events from SM backgrounds.
Drell-Yan production of dielectron pairs accounts for $150.1\pm14.1$ 
of these events, where we analyze Drell-Yan samples generated 
with $\pt(Z^0/\gamma^{*})>5\gevc$ \cite{njet} and normalize this production
to CDF data \cite{cdfdy} before applying the two jet requirement. 
There is also good agreement in the $\sum\et$ distributions for the remaining OS 
events and for the expected OS background (Figure ~\ref{fig:osdata}).

The sources of systematic uncertainty on the kinematic acceptances
for these analyses include 
initial and final state gluon radiation (ISR and FSR) 
(4\% for the charm squark analysis, 
$4-14$\% for the neutralino analysis),
uncertainty on the integrated luminosity (7\%),
electron identification (3\%),
structure functions (3\%),
Monte Carlo statistics ($1-5$\%),
jet energy scale (1\%),
and uncertainty on the trigger efficiency (1\%).
The total systematic uncertainty on the kinematic acceptance
is 10\% for the charm squark analysis,
while for the neutralino analysis it ranges from 10\% to 16\%.

We set limits on the cross section times branching ratio for the
two processes under study.  In each case we exclude
$
\sigma  \cdot Br \geq N_{95\%}/(A \cdot \epsilon_{trig} \cdot
\int {\cal L} \; dt),
$
where  $N_{95\%}$ is the
Poisson 95\% confidence level (C.L.) upper limit for observing 
zero events combined with a Gaussian distribution for the 
systematic uncertainty.  
For both analyses, $N_{95\%} = 3.1$ events.
The acceptance, $A$, is the product of the kinematic and 
geometric acceptance and
the efficiency of identifying two electrons and two jets, and
$\epsilon_{trig}$ is the trigger efficiency for dielectrons.
The integrated luminosity is $\int {\cal L} \; dt = 107 \pm 7$ \ipb.  

For the charm squark analysis,
$A$ is a very weak function of \mgluino\ and
ranges from 16.0\% to 16.6\%.  
For dielectrons with $\et(e)>15\gev$, $\epsilon_{trig}=98.4\%\pm1.3\%$.
We exclude $\sigma \cdot Br \geq 0.18$ pb independently of \mgluino.
Figure~\ref{fig:gg} shows the results for the charm squark analysis in
the gluino-squark mass plane.  Exclusion contours at the 95\% C.L.
are shown for two
values of the branching ratio $Br(\csquarkl \to ed)$, where we
compare our results to the next-to-leading order (NLO) $\gluino\gluino$ 
production cross
section \cite{nlo_gg} multiplied by the branching ratio to LS $ee$ from
Reference~\cite{choud_prd}. 
Our sensitivity vanishes for $M(\squark) \lesssim 260 \gevcc$.
In this region $\tilde{b}_L$ is lighter than $200 \gevcc$ 
(and thus lighter than \csquarkl)
due to the large top quark mass \cite{choud_prd},
so the decay of $\gluino \to \bar{b}\tilde{b}_L$ dominates and
$\gluino \to \bar{c} \tilde{c}_L$ is suppressed.  Since our analysis assumes
a non-zero \rpv\ coupling only for \csquarkl, the signal of LS
electrons with no \met\ disappears in this region of parameter space.

For the neutralino analysis, 
$A$ is determined for each squark and neutralino
mass pair and ranges from $3.7\%$ to $15.2\%$.
In this case, $\epsilon_{trig} = 96.5\%\pm1.9\%$, 
which is slightly lower than for
the charm squark analysis because the \et\ spectrum of the second
electron in the neutralino analysis is softer.  
We calculate the upper limit on the cross
section times branching ratio to LS \ee\ for each
squark and neutralino mass combination, and obtain $\sigma\cdot
Br$ limits which range as a function of the squark mass
from 0.81 pb to 0.26 pb for a light neutralino, and
from 0.35 pb to 0.20 pb for a heavy neutralino.
Figure~\ref{fig:lvlim} shows the results for the neutralino
analysis for the case of \ttbone\ production.  
Plotted are our 95\% C.L. upper limits 
along with the NLO cross section \cite{nlo_tt}
multiplied by the branching ratio to LS \ee.  
The branching ratio $Br(\stopo \to
c\,\chizero)$ is taken to be 1.0 \cite{stopclsp}.
We also assume $Br(\chizero\to
q\qbar' e)=Br(\chizero\to q\qbar' \nu)=1/2$, although the actual branching 
ratios are a function of the SUSY parameters \cite{dreiner_n1}.
Since each neutralino decays to $e^+$ or $e^-$ with equal probability, the
branching ratio to LS \ee\ is 1/8.  
The limit is shown for two extremes of the neutralino
mass, and excludes $\mstopo$ below
120 (135) \gevcc\ for a light (heavy) neutralino.
Similarly, the results for the case of five degenerate \ssb\ production are
displayed in Figure~\ref{fig:lvlim}.
In this case, the NLO cross section \cite{nlo_qq}
includes a gluino mass dependent $t$-channel contribution, and we
assume the branching ratio $Br(\squark \to q\,\chizero)=1.0$.
Thus, we set gluino and neutralino mass-dependent lower limits on the
degenerate squark mass in the range from 200 to 260 \gevcc.
The neutralino analysis presented here assumes that the only
non-zero \rpv\ coupling is \lamp. Since our analysis does not distinguish
the quark flavors in jet reconstruction, however,
the results are equally valid for any $\lambda^\prime_{1jk}$ coupling,
for which $j$ is 1 or 2 and $k$ is 1, 2 or 3.
  
We note that our limit for the neutralino decay analysis with 5 degenerate
squark flavors assumes the branching ratio $Br(\squark \to q\,\chizero) = 1.0$, 
whereas the branching ratio $Br(\csquarkl \to e\,d)$ must be
appreciable to explain the HERA results.  
However, even allowing for $Br(\squark \to q\,\chizero) < 1$, our analysis is
sensitive to the interesting region of 200 \gev, depending on 
$M(\gluino)$: for example, we can exclude the \rpv\ scenario with
$Br(\squark \to q\,\chizero) > 0.43$ for $M(\gluino) = 200 \gev$. 
For heavier gluino mass, the exclusion becomes weaker.

In conclusion, we find no evidence for LS dielectron plus multijet
events in 1.8 TeV \ppbar\ collisions
and set $\sigma \cdot Br$ limits on two \rpv\ SUSY processes
that could lead to this signature.
In the charm squark
analysis we exclude the scenario of $\mcsl=200 \gevcc$ as a
function of \mgluino\ and \msquark.  In the neutralino analysis
we set mass limits of $\mstopo>135\gevcc$ for a 
heavy neutralino ($\mchio=\mstopo-M(c)$)
and, for the degenerate squark, $\msquark>260\gevcc$ for a heavy
neutralino ($\mchio=\msquark-M(q)$) and a light gluino
(\mgluino=200\gevcc).

We thank the Fermilab staff and the technical staffs of the
participating institutions for their vital contributions.  
We also thank Debajyoti Choudhury, Sreerup Raychaudhuri, and Herbi
Dreiner for stimulating discussions. 
This work was
supported by the U.S. Department of Energy and National Science Foundation;
the Italian Istituto Nazionale di Fisica Nucleare; the Ministry of Education,
Science and Culture of Japan; the Natural Sciences and Engineering Research
Council of Canada; the National Science Council of the Republic of China; 
the Swiss National Science Foundation; and the A. P. Sloan Foundation.

\def\Journal#1#2#3#4{{#1} {\bf #2}, #3 (#4)}
\def\NCA{Nuovo Cimento}
\def\NIM{Nucl. Instrum. Methods}
\def\NIMA{{Nucl. Instrum. Methods} A}
\def\NPB{{Nucl. Phys.} B}
\def\PLB{{Phys. Lett.}  B}
\def\PRL{Phys. Rev. Lett.}
\def\PRD{{Phys. Rev.} D}
 \def\PR{Phys. Rep.}
\def\ZPC{{Z. Phys.} C}
\def\MPL{{Mod. Phys. Lett.} A}

\renewcommand{\baselinestretch}{1}

\begin{figure}
\begin{center}
\epsfig{file=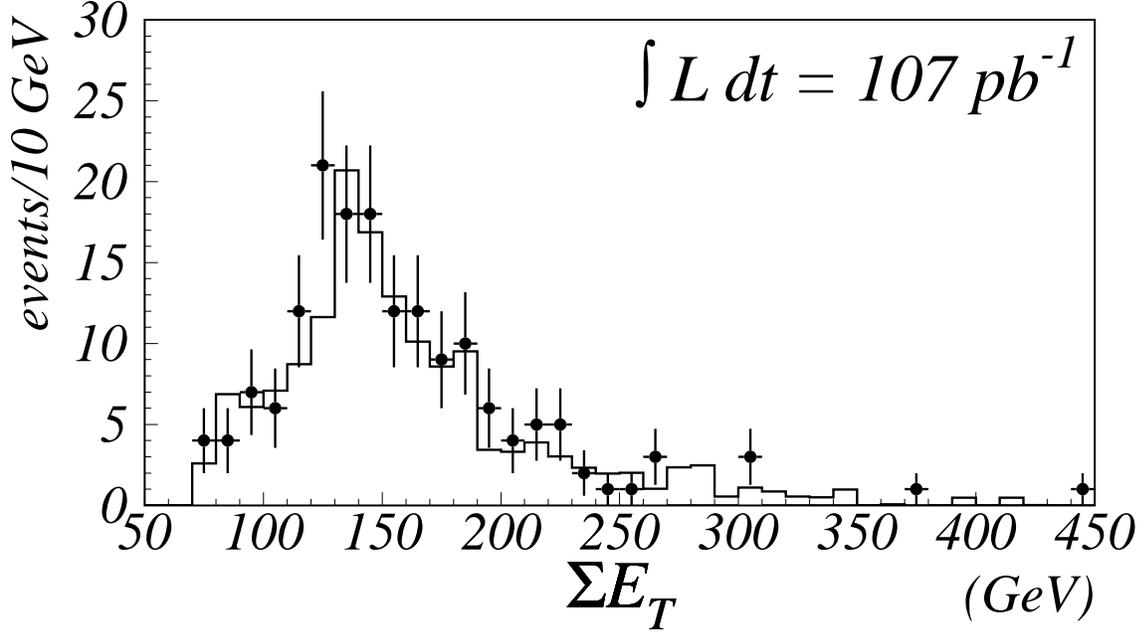,width=\textwidth} 
\caption{Scalar sum of transverse energy for the two electrons
and two leading jets for the remaining 165 OS events after all 
selection (points) and expected background from SM processes 
(histogram).  These events are dominated by Drell-Yan $\epem$ pairs
plus two or more jets.}
\label{fig:osdata}
\end{center}
\end{figure}
\begin{figure}
\begin{center}
\epsfig{file=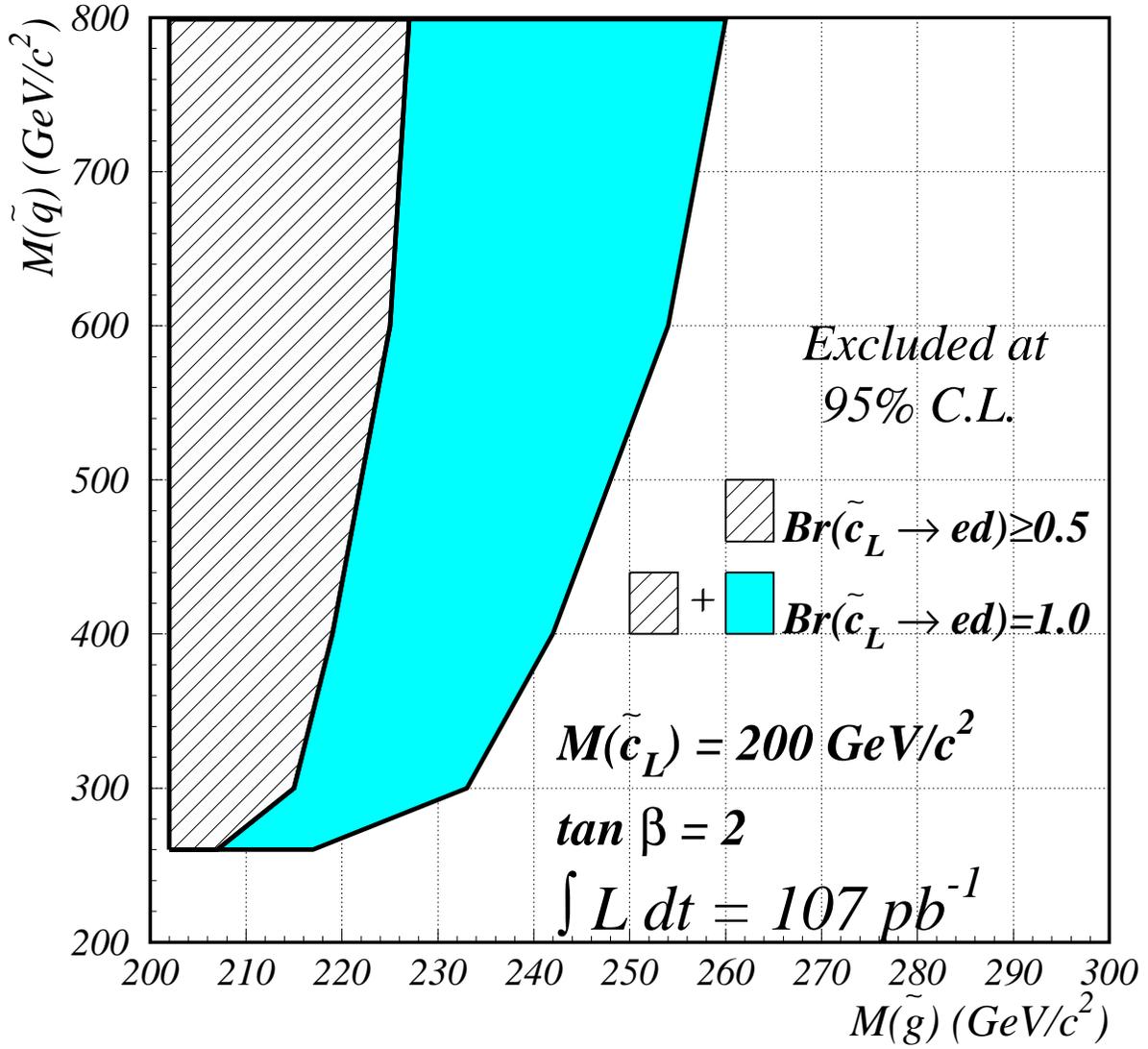,width=\textwidth} 
\caption{Exclusion region in the \gluino\ - \squark\ mass plane 
for the charm squark analysis.  The branching ratio 
to LS \ee\ is calculated using the scenario in
Reference~\protect\cite{choud_prd}, which requires $\mgluino>\mcsl$.}
\label{fig:gg}
\end{center}
\end{figure}
\begin{figure}[h]
\begin{center}
\epsfig{file=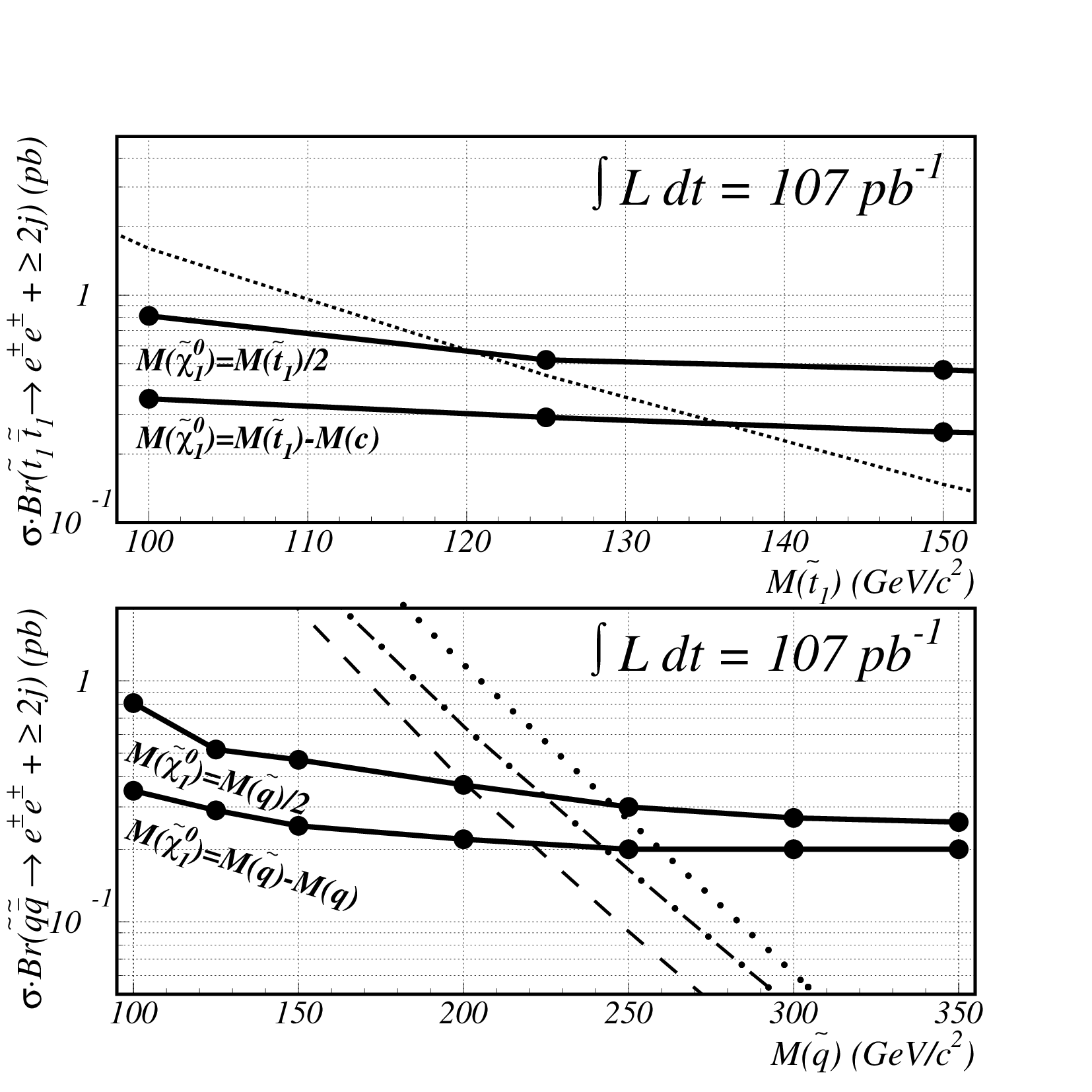,width=\textwidth} 
\caption{Top: upper limits on the cross section times branching ratio
for \ttbone\ production decaying to electrons and jets
via neutralinos (solid lines).
The dashed curve is the theoretical prediction for $\sigma \cdot Br$.
Bottom: upper limits on the cross section times branching ratio
for the production of 5 degenerate squark flavors decaying to 
electrons and jets via neutralinos (solid lines).
Also shown is the theoretical prediction for $\sigma \cdot Br$
for three values of the gluino
mass: 200 \gevcc\ (dotted line), 500 \gevcc\ (dot-dashed line), and 1
\tevcc\ (dashed line).}
\label{fig:lvlim}
\end{center}
\end{figure}


\begin{thebibliography}{99}
\bibitem{njet} We use Monte Carlo samples generated in this
	transverse momentum range because they reproduce better
	the jet multiplicities seen in our data.
\end{thebibliography}
\end{document}